\documentclass[pra,aps,twocolumn,showpacs,showkeys]{revtex4}
\usepackage{graphicx}
\usepackage{amsmath}

\begin{document}

\title{Transmission of optical coherent state qubits}
\author{S. Glancy}
\email{sglancy@boulder.nist.gov}
\author{H. M. Vasconcelos}
\email{Hilma.M.DeVasconcelos.2@nd.edu}
\affiliation{Department of Physics, University of Notre Dame, Notre Dame, Indiana 46556, USA}
\author{T. C. Ralph}
\email{ralph@physics.uq.edu.au}
\affiliation{Centre for Quantum Computer Technology, Department of Physics, University of Queensland Brisbane, QLD 4072, Australia}
\date{\today}
\begin{abstract}
We discuss the long distance transmission of qubits encoded in optical coherent states. Through absorption these qubits suffer from two main types of errors, the reduction of the amplitude of the coherent states and accidental application of the Pauli $Z$ operator. We show how these errors can be fixed using techniques of teleportation and error correcting codes.
\end{abstract}
\maketitle

\section{Introduction}
Quantum optics has been a fertile field for experimental tests of quantum information science. We expect that optical methods will be especially useful in applications that require the communication of quantum information over long distances. This has already been proven to be the case for quantum encryption.  Most research in this field has focused on the communication of single photon qubits. In this case the dominant source of decoherence is photon absorption, so errors can usually be avoided by engineering experiments that use some kind of coincidence counting. Then, either all expected photons are observed, and no errors have occurred; or some photons are absorbed, fewer photons are observed than expected, and the experiment must be repeated.

In this paper we will discuss the communication of qubits that are encoded using multi-photon optical coherent states \cite{Cochrane, Jeong, Enk, Ralph01, Ralph03}.  Because these are multiphoton states, they are more robust against small levels of absorption. However, the coherent state qubits exhibit their own errors, so we will later show how to correct these errors. To perform our error correcting procedure, we will need a resource of ``cat states'' of the form $|-\alpha\rangle+|\alpha\rangle$, beam splitters, and high efficiency photon counters.

The coherent state $|\alpha\rangle$ is defined to be the eigenstate of the annihilation operator $\hat{a}$ with eigenvalue $\alpha$, which may be any complex number.
\begin{equation}
\hat{a}|\alpha\rangle=\alpha|\alpha\rangle.
\end{equation}
In the Fock basis the coherent state has the decomposition
\begin{equation}
|\alpha\rangle=e^{-|\alpha|^2/2}\sum_{n=0}^{\infty}\frac{\alpha^n}{\sqrt{n!}}|n\rangle,
\end{equation}
and it is generated from the vacuum, $|0\rangle$, by the displacement operator $D(\alpha)=\exp(\alpha \hat{a}^{\dagger}-\alpha^{\ast} \hat{a})$. For a more thorough introduction to coherent states see \cite{Mandel} for example.
Specifically we will consider two different methods of encodings qubits using coherent states.  The first of these is the $(-,+)$ encoding, where $|0 \rangle_{L} \equiv |-\alpha \rangle$, and $|1 \rangle_{L} \equiv |\alpha \rangle$. An arbitrary qubit appears as
\begin{equation}
|Q_{\pm}(\alpha)\rangle = \frac{1}{\sqrt{N(\alpha)}}(\mu |-\alpha\rangle +\nu |\alpha\rangle),
\end{equation}
where $\mu$ and $\nu$ are complex numbers satisfying $|\mu|^{2}+|\nu|^{2}=1$. To simplify the notation, throughout the remainder of this paper, we assume $\alpha$ is a real number. $N(\alpha)=1+e^{-2\alpha^2}(\mu\nu^{\ast}+\mu^{\ast}\nu)$ and is a normalization factor, made necessary because the states $|-\alpha\rangle$ and $|\alpha\rangle$ are not exactly orthogonal, although for a sufficiently large $\alpha$ (say $\alpha \ge 2$) they are very nearly orthogonal ($N \approx 1$). In the second encoding we will consider $|0 \rangle_{L} \equiv |0 \rangle$, and $|1 \rangle_{L} \equiv |2 \alpha \rangle$. Using this $(0,\alpha)$ encoding, we will represent an arbitrary qubit with
\begin{equation}
|Q_{0\alpha}(\alpha)\rangle = \frac{1}{\sqrt{N(\alpha)}}(\mu |0\rangle+\nu |2\alpha\rangle).
\end{equation}
Both of these encoding require the same normalization factor, and we can easily translate from one encoding to the other by using the displacement operator $D(\pm\alpha)$. Methods for performing universal logic operations on these qubits have already been discussed in \cite{Ralph01, Ralph03}, but we will briefly review the operations needed to correct the decoherence caused during long distance transmission.

\section{Decoherence}

Our first task is to characterize the type of errors that will effect the qubits as they travel through a long optical fiber. We assume photon loss is the dominant decoherence mechanism. This can be modeled by assuming some of the field is lost in transit via a beam splitter type interaction. The qubit enters one mode of the beam splitter, and the vacuum enters the other mode. After passing through the beam splitter some of the qubit's energy and information will be transfered to the second mode of the beam splitter and lost to the environment. Only a single beam splitter is necessary to characterize the effects of any number of photon loss events and mechanisms. Even though the photons will actually be lost to a myriad of different modes, our assumption that all photons are lost to a single mode is sufficient, because we must trace over this mode (or these modes) to calculate the state of the transmitted qubit.

We will characterize the optical fiber used to transmit the qubit as having an exponential energy loss $e^{-\lambda L}$, where $\lambda$ is the loss coefficient of the fiber, and $L$ is the transmission distance. High grade commercial fibers typically have $\lambda\sim 0.06/\text{km}$ \cite{Wu}. We will chose a beam splitter transmissivity of $\eta=e^{-\lambda L}$ to study the decoherence behavior of both qubit encodings. After transmission the state of the qubit and the loss mode (denoted with an $l$) becomes 
\begin{eqnarray}
|Q_{\pm} \rangle_{T} & = & \mu |-\alpha\sqrt{\eta}\rangle |-\alpha\sqrt{1-{\eta}}\rangle_{l} \nonumber \\
& & + \nu |\alpha\sqrt{\eta}\rangle|\alpha\sqrt{1-\eta}\rangle_{l}
\label{pm1}
\end{eqnarray}
or
\begin{equation}
|Q_{0\alpha}\rangle_{T} = \mu |0\rangle|0\rangle_{l} + \nu |2\alpha\sqrt{\eta}\rangle|2\alpha\sqrt{1-\eta}\rangle_{l}.
\label{0alpha1}
\end{equation} 

One question which is natural to address at this point is ``Which encoding method results in the greatest loss of information from the qubit mode?'' It is tempting to think, because of the energy difference between the logical states in the $(0, \alpha$) encoding, that this encoding will suffer a catastrophic ``spontaneous emission'' type error under photon loss. However this is not correct as can be seen from the following argument.

The state of the qubit mode after transmission is found by performing a trace over the loss mode. Therefore we may perform any unitary operation on the loss mode, because that would just be equivalent to performing the trace using a different set of basis states. The amount of information in the qubit mode would be preserved under any unitary operation on that mode as well.  One can easily see that if we apply $D(\alpha\sqrt{\eta})$ to the qubit mode and $D(\alpha\sqrt{1-\eta})$ to the loss mode in Eq.~(\ref{pm1}) we obtain a state equal to that in Eq.~(\ref{0alpha1}). By this argument we expect that both encodings should exhibit the same amount of decoherence. Learning exactly how this decoherence is manifest will require closer inspection.

We must find the final state of the qubit after transmission. Because information was lost to the environment during transmission, the qubit is in a mixed state, and must be described by a density operator.  We first construct a density operator for the qubit and loss modes, which is given by $|Q_{\pm}\rangle_{T} {}_{T}\langle Q_{\pm}|$ in the $(-,+)$ encoding. A partial trace over the loss mode $l$ is performed by calculating
\begin{equation}
\rho_{\pm}=\sum_{n=0}^{\infty} {}_{l}\langle n|Q_{\pm}\rangle_{T} {}_{T}\langle Q_{\pm}|n\rangle_{l}
\end{equation}
Without the need for any approximation we obtain \cite{Cochrane}
\begin{eqnarray}
\rho_{\pm} & = & \left(1-P_e\right)\left(\mu |- \alpha\sqrt{\eta}\rangle + \nu | \alpha\sqrt{\eta}\rangle\right) \times \text{h.c.} \nonumber \\
& & + P_e\left(\mu |- \alpha\sqrt{\eta}\rangle - \nu |\alpha\sqrt{\eta}\rangle\right) \times \text{h.c.},
\end{eqnarray}
which we can rewrite as
\begin{eqnarray}
\rho_{\pm} & = & \left(1-P_e\right)|Q_{\pm}(\alpha\sqrt{\eta})\rangle\langle Q_{\pm}(\alpha\sqrt{\eta})| \nonumber \\
& & + P_e Z |Q_{\pm}(\alpha\sqrt{\eta})\rangle\langle Q_{\pm}(\alpha\sqrt{\eta})| Z,
\end{eqnarray}
The calculation for the $(0,\alpha)$ encoding is similar and yields
\begin{eqnarray}
\rho_{0\alpha} & = & \left(1-P_e\right)\left(\mu |0\rangle + \nu | 2\sqrt{\eta} \alpha \rangle\right) \times \text{h.c.} \nonumber \\
& & + P_e\left(\mu |0\rangle - \nu | 2\sqrt{\eta} \alpha \rangle\right) \times \text{h.c.},
\end{eqnarray}
which is equivalent to
\begin{eqnarray}
\rho_{0\alpha} & = & \left(1-P_e\right) |Q_{0\alpha}(\alpha\sqrt{\eta})\rangle \langle Q_{0\alpha}(\alpha\sqrt{\eta})| \nonumber \\
& & + P_e Z |Q_{0\alpha}(\alpha\sqrt{\eta})\rangle \langle Q_{0\alpha}(\alpha\sqrt{\eta})| Z,
\end{eqnarray}
where $P_e=\frac{1}{2}\left(1-e^{-2(1-\eta)\alpha^{2}}\right)$, $\text{h.c.}$  is the hermitian conjugate of the previous factor, and $Z$ is the Pauli $Z$-operator defined by $Z(\mu |0 \rangle_{L} + \nu |1 \rangle_{L}) = (\mu |0 \rangle_{L} - \nu |1\rangle_{L})$. From these calculations one can see that the decoherence of the qubits in both encodings is manifest in two ways. First the amplitude of the coherent states is changed from $\alpha$ to $\alpha\sqrt{\eta}$. Second with probability $P_e$, the Pauli $Z$ operator is applied, producing a phase flip in the qubit basis.

\section{Basic Operations}

Before we begin our discussion of how this decoherence can be corrected, we will first describe how to perform a few fundamental logic operations on the coherent state qubits. This will not be a universal set of logic gates but only the tools we need for error correction.

{\em Displacement operator}. The displacement operator can be simulated by sending a qubit into a highly transmissive beam splitter along with a high amplitude coherent state. The beam splitter performs the unitary transformation
\begin{equation}
U_{BS}(\theta)=e^{\theta\left(\hat{a_1}\hat{a_2}^{\dagger}-\hat{a_1}^{\dagger}\hat{a_2}\right)},
\end{equation}
where $\hat{a_1}$ and $\hat{a_2}$ are the annihilation operators for the two modes entering the beam splitter, and the transmissivity is given by $\cos^2\theta$. For example, suppose the qubit modes is in state $|\alpha\rangle$, and we send it and a second mode in the state $\beta$ into a beam splitter with transmissivity $\eta$. This produces
\begin{equation}
|\alpha\rangle|\beta\rangle \rightarrow |\sqrt{\eta}\alpha-\sqrt{1-\eta}\beta\rangle|\sqrt{\eta}\beta+\sqrt{1-\eta}\alpha\rangle
\end{equation}
In the limit that $\beta \rightarrow \infty$, $\eta \rightarrow 1$, and $-\beta\sqrt{1-\eta} \rightarrow \gamma$, this interactions produces $D(\gamma)$ on the qubit. This can be seen by performing a trace over the second mode in the above expression and taking the appropriate limits. Because $|\beta\rangle$ is just a coherent state ({\em not} a superposition of two coherent states) it is not difficult to approximate the limit $\beta \rightarrow \infty$. We can now use this technique to translate between the two qubit encodings with negligible decoherence.

{\em Bit-flip gate.} The bit-flip gate performs the operation $|-\alpha\rangle \rightarrow |\alpha\rangle$ and $|\alpha\rangle \rightarrow |-\alpha\rangle$ in the $(-,+)$ encoding. It is equivalent to the logical NOT operation or the Pauli $X$ matrix in the logical qubit basis. To achieve this transformation we simply apply the Hamiltonian for free space evolution $H=\hbar\omega\hat{a}^{\dagger}\hat{a}$ for a time $t=\frac{\pi}{\omega}$, where $\omega$ is the frequency of the light. This produces the unitary operator
\begin{equation}
U(\pi)=e^{i\pi \hat{a}^{\dagger}\hat{a}}
\end{equation}
We can easily apply this Hamiltonian by delaying the flight of the qubit by one half cycle with respect to the local oscillator field. Note that we can perform $X$ for qubits in the $(0,\alpha)$ basis with the combination $U(\pi)D(-2\alpha).$

{\em Bell state preparation.} Many of the operations of a coherent state quantum computer, especially those based on quantum teleportation, require the use of Bell states of the form
\begin{equation}
|B_{\pm}(\alpha)\rangle=\frac{1}{\sqrt{2+2e^{-2 \alpha^2}}}(|-\alpha\rangle|-\alpha\rangle + |\alpha\rangle|\alpha\rangle).
\end{equation}
The preparation of these Bell states is a nontrivial matter. A Bell state can be made by creating a ``cat state'' of the form $|-\sqrt{2}\alpha\rangle+|\sqrt{2}\alpha\rangle$, and sending the cat state and the vacuum into the two input ports of a 50/50 beam splitter $U_{BS}(\pi/4)$. Then the beam splitter's output ports will contain the two entangled modes of the Bell state.

Therefore the problem of generating Bell states is reduced to the construction of a source of cat states. Cat states can be produced by sending a coherent state into a nonlinear medium exhibiting the Kerr effect\cite{Yurke}. Although sufficiently large Kerr nonlinearities have been difficult to produce, significant progress is being made in this area. For a discussion of an experiment demonstrating the observation of the Kerr effect see \cite{Kang}.

It is also possible to produce approximate cat states using a squeezing interaction, linear optical devices, and photon counters \cite{Ralph03,Song,Dakna}. The success of these methods depends on detecting a particular number of photons in the photon counter, so for a single attempt there is a small probability to produce a high fidelity cat state. Ralph {\it et al.} calculate that a cat state with a fidelity of $0.95$ can be produced with a probability greater than $1\%$. This presents the technical problem of constructing a system which can rapidly repeat cat state production attempts (or can simultaneously perform many attempts), so that cat states can be had on demand. The greatest challenge to this cat state production technique is that it relies on very high efficiency photon detectors which can distinguish between for example 5, 6, or 7 photons. While such photon detectors do not yet exist, this is also a very promising field of research. For a demonstration of a high efficiency photon counter see \cite{Waks}. A promising new method to produce cat states was recently proposed by Lund {\it et al.} in \cite{Lund}. They explain how high amplitude cat states can be produced from squeezed single photon states, using beam splitters and inefficient photon detectors. This scheme also has a large probability of failure during a given attempt to produce a cat, but it can be implemented with current photon detectors. It appears likely that the production of cat states (and Bell states) is a goal that is achievable in the near future.

{\em Teleportation and Z.} Quantum teleportation is a surprisingly straightforward procedure for coherent state qubits \cite{Jeong,Enk}. It is an essential tool for performing many logic operations on coherent state qubits, and we will later show how it can be used to correct the amplitude lost during qubit transmission.

\begin{figure}
\begin{center}
\includegraphics[width=80mm]{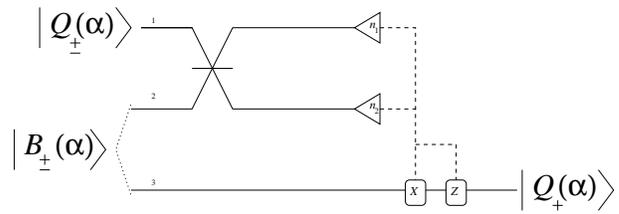}
\end{center}
\caption{\label{teleport1}Schematic diagram for teleportation of a qubit in the $(-,+)$ encoding. The beam splitter has a transmissivity of $1/2$. The photon detectors register $n_1$ and $n_2$ photons, and at least one of $n_1$ and $n_2$ must equal zero. If $n_1\neq 0$, we apply $X$. If an odd number of photons is detected, we apply $Z$. If both $n_1=n_2=0$, then the teleportation fails.}
\end{figure}

To perform the teleportation in the $(-,+)$ encoding, we need to mix the qubit $|Q_{\pm}(\alpha)\rangle$, with one of the Bell state's two modes in a 50/50 beam splitter. The schematic diagram for the teleportation is shown in the Fig.~\ref{teleport1}.  

The state of the three modes after the qubit and the Bell state mix in the beam splitter is given by
\begin{eqnarray}
|T_{\pm}(\alpha)\rangle & = & \mu (|0\rangle_{1} |\sqrt{2}\alpha\rangle_{2} |\alpha\rangle_{3}) \nonumber \\
& & + \mu (|-\sqrt{2}\alpha\rangle_{1} |0\rangle_{2} |-\alpha\rangle_{3}) \nonumber \\
& & + \nu (|0\rangle_{1} |-\sqrt{2}\alpha\rangle_{2} |-\alpha\rangle_{3}) \nonumber \\
& & + \nu (|\sqrt{2}\alpha\rangle_{1} |0\rangle_{2} |\alpha\rangle_{3}),
\label{tpm}
\end{eqnarray}
After passing through the beam splitter, modes $1$ and $2$ are sent to photon counters, which register $n_{1}$ and $n_{2}$ photons. Depending on the results of the measurement, the qubit may need to be corrected. We can see from Eq.~(\ref{tpm}) that it is impossible for both detectors to register photons. If $n_1$ is an even number, and $n_2=0$, then the qubit needs no correction. If $n_1$ is an odd number, and $n_2=0$, then $Z$ must be applied to the qubit. If $n_1=0$, and $n_2$ is an even number, then $X$ must be applied to the qubit. If $n_1=0$, and $n_2$ is an odd number, then both $X$ and $Z$ must be applied. There is also a small probability $\sim e^{-\alpha^2}$ for measuring zero photons in both photon counters, indicating a failure of the teleportation. This occurs because our ``basis'' states are not orthogonal.

\begin{figure}
\begin{center}
\includegraphics[width=80mm]{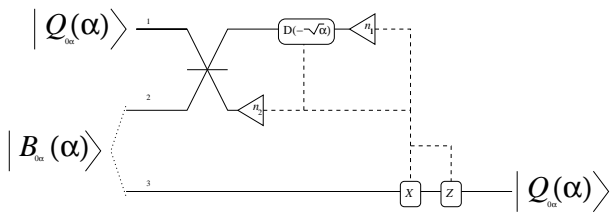}
\end{center}
\caption{\label{teleport2}Schematic diagram for teleportation of a qubit in the $(0,\alpha)$ encoding. The beam splitter has a transmissivity of $1/2$. If $n_2=0$, then we must apply $D(-\sqrt{2}\alpha)$ to the first mode, and if $n_1$ is odd we apply $Z$ to the third mode. If $n_2\neq 0$, we apply $X$, and if $n_2$ is odd we also apply $Z$. If both $n_1=n_2=0$ the teleportation fails.}
\end{figure}

For the $(0,\alpha)$ encoding, the teleportation scheme is very similar to the scheme used in the $(-,+)$ encoding, as we can see in Fig.~\ref{teleport2}. The basic difference is the displacement operation done to the first mode after mixing the qubit and the Bell state in a 50/50 beam splitter. 

The input state in this case is $|Q_{0 \alpha}(\alpha)\rangle$ and the Bell state is
\begin{equation}
|B_{0\alpha}(\alpha)\rangle=\frac{1}{\sqrt{2+2 e^{- 4 \alpha^2}}}(|0\rangle_{2}|0\rangle_{3} + |2 
\alpha\rangle_{2}|2 \alpha\rangle_{3})).
\end{equation}
The state after the beam splitter is given by
\begin{eqnarray}
|T_{0 \alpha}(\alpha)\rangle & = & \mu (|0\rangle_{1} |0\rangle_{2} |0\rangle_{3}) \nonumber \\
& & + \mu (|\alpha/\sqrt{2}\rangle_{1} |\alpha/\sqrt{2}\rangle_{2} |\alpha\rangle_{3}) \nonumber \\
& & + \nu (|\alpha/\sqrt{2}\rangle_{1} |-\alpha/\sqrt{2}\rangle_{2} |0\rangle_{3}) \nonumber \\
& & + \nu
(|\sqrt{2}\alpha\rangle_{1} |0\rangle_{2} |\alpha\rangle_{3}).
\end{eqnarray}
After the beam splitter, we first measure $n_2$ the number of photons in mode $2$. If $n_2$ is an even number greater than zero, then mode 1 must contain the vacuum, and the qubit must be corrected with $X$. If $n_2$ is an odd number then the qubit must be corrected with both $X$ and $Z$. However, if $n_2=0$, then we must apply the displacement $D(\sqrt{2}\alpha)$ to mode 1 and count the number of photons in this mode. If $n_1$ is even, the qubit needs no correction. If $n_1$ is odd, the qubit must be corrected with $Z$. 

\begin{figure}
\begin{center}
\includegraphics[bb = 90 578 319 716, clip, width=80mm]{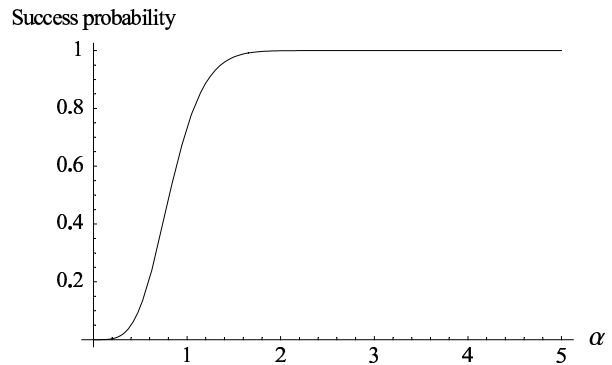}
\end{center}
\caption{\label{teleportsuccess}Teleportation success probability versus $\alpha$ for $\mu=\nu=\frac{1}{\sqrt{2}}$}
\end{figure}

The probability that the teleportation fails because we detect zero photons in both photon counters decreases rapidly as $\alpha$ increases. In Fig.~\ref{teleportsuccess} we plot the probability to successfully perform a teleportation, $P_{s}=1-|{}_{2}\langle 0| {}_{1}\langle 0| T_{\pm}(\alpha)\rangle|^2 = 1-|{}_{2}\langle 0|{}_{1}\langle 0| D(-\sqrt{2}\alpha)T_{0\alpha}(\alpha)\rangle|^2$ as a function of $\alpha$. The two encodings give the same success probability for any choice of $\mu$ and $\nu$.

Notice that we have not yet explained how to accomplish $Z$, which is required whenever an odd number of photons is detected during the Bell basis measurement. This is because we propose to use the teleportation operation itself when $Z$ is needed. When an odd number of photons is detected, signaling that the teleportation has resulted in the state $Z|Q\rangle$ we simply attempt teleportation again and hope to again detect an odd number. Because the photon detectors always signal when $Z$ has been performed we are free to simply try again.

{\em Hadamard gate.} The Hadamard gate $H$ performs the transformation $|0\rangle_L \rightarrow |0\rangle_L+|1\rangle_L$ and $|1\rangle_L \rightarrow |0\rangle_L - |1\rangle_L$ (neglecting normalization). It is easiest to perform in the $(0,\alpha)$ encoding, in which the steps required to implement $H$ are similar to those needed for teleportation. We first put our qubit $|Q_{0\alpha}(\alpha)\rangle$ and one half of the Bell state $|B_{0\alpha}(\alpha)\rangle$ into the beam splitter described by the interaction
\begin{equation}
U_{iBS}(\theta)=e^{i\theta\left(\hat{a_1}\hat{a_2}^{\dagger}+\hat{a_1}^{\dagger}\hat{a_2}\right)},
\end{equation}
where $\hat{a_1}$ and $\hat{a_2}$ are the annihilation operators for the two modes entering the beam splitter. We choose $\theta=\pi/2\alpha^2$. Using the approximation that $\alpha \rightarrow \infty$, this interaction becomes approximately equal to a controlled sign flip in the logical qubit basis, which performs the identity to all input qubit basis states except $|1\rangle_{L}|1\rangle_{L}$ which becomes $-|1\rangle_{L}|1\rangle_{L}$.

We then need to measure the two modes exiting the beam splitter in the basis whose basis states are eigenstates of $X$. This can be accomplished by applying the displacement operator $D(-\alpha)$ to each of these modes, converting them to the $(-,+)$ encoding. In this encoding the eigenstates of $X$ are the cat states $|-\alpha\rangle \pm |\alpha\rangle$, which are distinguishable because the $+$ cat has only even numbers of photons and the $-$ cat has odd numbers of photons. We can therefore measure in the $X$ eigenbasis by detecting either even or odd numbers of photons. The remaining qubit must then be corrected using $X$ and $Z$ (depending on the measurements) to produce $H|Q_{0\alpha}(\alpha)\rangle$.

This procedure for executing $H$ has a close analogy with the teleportation of logic gates as described by Gottesman and Chuang in \cite{Gottesman}. They show how logic operations may be performed on qubits by executing a teleportation like procedure using modified Bell states. In our implementation of the Hadamard gate we use a standard Bell state, but we perform the joint measurement between the qubit and one qubit from the Bell state in a different basis.

\begin{figure}
\begin{center}
\includegraphics[width=80mm]{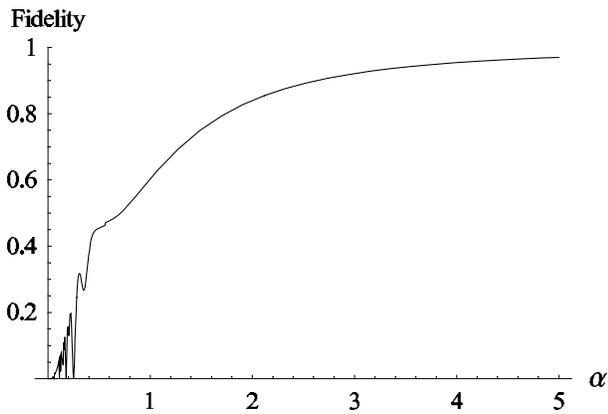}
\end{center}
\caption{\label{hfidelity}The fidelity of the Hadamard transformation as a function of coherent state amplitude $\alpha$. Here we use the worst case input qubit $|Q\rangle=|2\alpha\rangle$.}
\end{figure}

To characterize the effectiveness of this gate we will examine its fidelity when $\alpha$ is small. Suppose that the two detectors count $n_a$ and $n_b$ photons. We will then call the resulting qubit $|HQ_{n_{a},n_{b}}\rangle$, which we hope is approximately equal to $H|Q\rangle$. On average the procedure to implement $H$ will result in the mixed state
\begin{equation}
\rho=\sum^{\infty}_{n_{a}=0}\sum^{\infty}_{n_{b}=0} P(n_{a},n_{b}) |HQ_{n_{a},n_{b}}\rangle \langle HQ_{n_{a},n_{b}}|,
\end{equation}
where $P(n_{a},n_{b})$ is the probability to detect the combination $n_a$ and $n_b$. The fidelity of this operation is given by
\begin{equation}
F=\langle Q|H \rho H|Q\rangle.
\end{equation}
We plot this fidelity as a function of $\alpha$ in Fig.~\ref{hfidelity}. There we use the worst case input qubit $|Q\rangle=|2\alpha\rangle=|1\rangle_{L}$.  The fluctuating structure for small $\alpha$ is caused by the oscillations of the beam splitter's transmissivity $\eta=\cos^{2}(\pi/2\alpha^{2})$.

In order to further improve the fidelity for small $\alpha$ we can choose to operate the Hadamard gate in a probabilistic manner. The combinations of $n_{a}$ and $n_{b}$ which yield lower fidelities will simply be rejected and classified as failures of the gate. Suppose we desire to operate the gate with a fidelity of 0.99, so we will exclude just enough $n_{a}$, $n_{b}$ combinations to ensure this fidelity. We would then like to know, ``What is the probability that this operation will succeed?''  With $\alpha=2$ and the worst case input qubit, the Hadamard gate will succeed with a probability of 0.29.  Using $\alpha=4$ the success probability is 0.59. It is possible to protect the qubits from destruction when the Hadamard gate fails by using a second layer of teleportation as described in \cite{Gottesman}. We would apply the Hadamard gate itself (meaning the application of $U_{iBS}$ followed by measurement in the eigenbasis of $X$) to one of the qubits of a Bell state. Then the information bearing qubit is ``teleported'' using this modified Bell state. The result of this ``teleportation'' will be equal to $H|Q\rangle$. Because the Bell state does not contain any information, when the Hadamard gate fails we can simply produce another Bell state and attempt to perform $H$ again.

While teleportation and the Hadamard gate present some significant technical challenges, they are not entirely beyond the horizon of current experiments. Using these tools, we will show how the decoherence caused to qubits during transmission can be corrected.

\section{$Z$ Error Correction}

\begin{figure}
\begin{center}
\includegraphics[width=80mm]{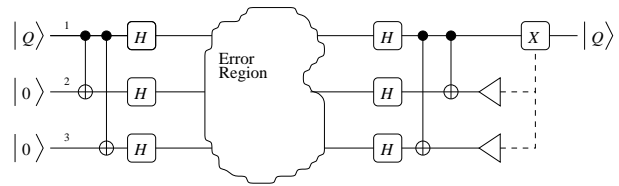}
\end{center}
\caption{\label{3bitecc}Quantum logic circuit for the standard three qubit code used to correct $Z$ errors. The CNOT gates are controlled by the qubit $|Q\rangle$ and apply the NOT to the qubits intersecting the open circles. The $H$'s represent Hadamard transformations. The diagram shows the encoding of the qubits, the qubits' passage through the decoherence region, and the decoding of the qubits. The triangles represent qubit measurements. If both measurements are $|0\rangle$ or if both are $|1\rangle$, then $|Q\rangle$ exits from the decoding.  If the measurements detect different results, then the qubit must be corrected with $Z$.}
\end{figure}

The erroneous application of the Pauli $Z$ operator can be corrected using a standard error correcting code \cite{Nielsen}, that encodes a single qubit onto three qubits. The quantum circuit that is traditionally used to accomplish this encoding is depicted in Fig.~\ref{3bitecc}. It requires two controlled not gates (CNOTs) and three Hadamard transformations in the encoding stage and another two CNOTs and three $H$'s in the decoding gates. The CNOT gate could be applied using the techniques of coherent linear optic quantum computing \cite{Ralph03}. However, these logic operations are quite difficult to accomplish. Rather than performing the encoding this way, we will describe how the properties of coherent state qubits actually allow for much more efficient method for encoding and decoding.

The effect of the first three CNOTs in Fig.~\ref{3bitecc} is to perform the transformation.
\begin{equation}
\left(\mu|0\rangle+\nu|1\rangle\right)|0\rangle|0\rangle \rightarrow \mu|0\rangle|0\rangle|0\rangle+\nu|1\rangle|1\rangle|1\rangle.
\label{CNOTs}
\end{equation}
At this point the qubit is protected from $X$ errors. After each qubit receives an $H$, they are protected from $Z$ errors.

\begin{figure*}
\begin{center}
\includegraphics[width=170mm]{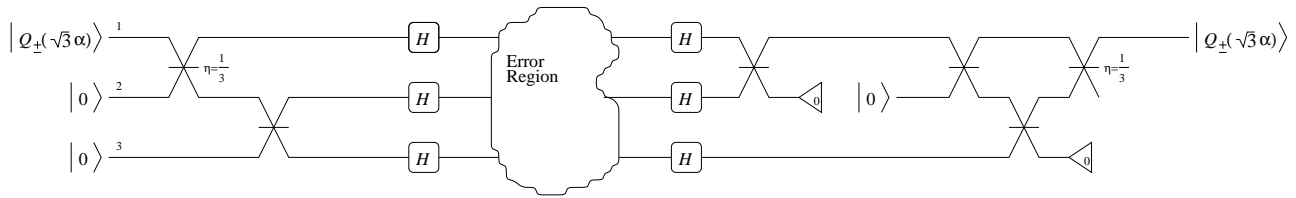}
\end{center}
\caption{\label{cscode}Implementation of the three qubit code on coherent state qubits in the $(-,+)$ encoding. The first and last beam splitters have a transmissivity of $1/3$, and the others have a transmissivity of $1/2$. The qubit $|Q_{\pm}\rangle$ will emerge from the first mode without error provided that both detectors register zero photons.}
\end{figure*}

The structure of coherent state qubits makes the transformation Eq.~(\ref{CNOTs}) surprisingly simple. In fact we can perform this transformation on an arbitrary, unknown qubit without the need for performing the complicated CNOT gates. Suppose we are given an unknown qubit, and we want to encode it against $Z$ errors before sending it on to another party across a great distance. Assume that the qubit arrives in the form $\mu|-\alpha\rangle+\nu|\alpha\rangle$, where $\alpha$ is known to us (It may be any complex number.), but $\mu$ and $\nu$ are not known. First it may be helpful (thought not necessary) to increase the amplitude of the coherent states used to encode this qubit. This can be accomplished by teleporting the qubit onto a Bell state, one of whose modes (the mode which is measured) has the amplitude $\alpha$ and the other mode (which will contain the teleported qubit) has the new amplitude $\sqrt{3}\alpha$. This allows us to perform the transformation
\begin{equation}
\mu|-\alpha\rangle + \nu|\alpha\rangle \rightarrow \mu|-\sqrt{3}\alpha\rangle + \nu|\sqrt{3}\rangle.
\end{equation}
The details of this teleportation procedure are described in the following section. We then append two modes, containing only vacuum states, to the qubit. The qubit and the two vacuum modes are sent through the circuit pictured in Fig.~\ref{cscode}.

The state $\left(\mu|-\sqrt{3}\alpha\rangle+\nu|\sqrt{3}\alpha\rangle\right)|0\rangle|0\rangle$ enters the circuit from the left. The first and second modes mix in a beam splitter whose transmissivity is $\eta=\frac{1}{3}$. Then the second and third modes mix at a beam splitter with transmissivity $\eta=\frac{1}{2}$. This prepares the state $\mu|-\alpha\rangle|-\alpha\rangle|-\alpha\rangle+\nu|\alpha\rangle|\alpha\rangle|\alpha\rangle$. We have therefore accomplished the transformation Eq.~(\ref{CNOTs}) without knowledge of the value of the qubit or the use of traditional CNOT operations. The Hadamard transformations may then be applied using the procedure described in the previous section.

After the protected qubits are transmitted down long fibers, they must be decoded, and an error syndrome must be measured. To decode we first apply an $H$ to each qubit. We can then simulate the two CNOTs used for the decoding in Fig.~\ref{3bitecc} with the decoding circuit pictured in Fig.~\ref{cscode}. Let us first examine the case in which no errors have occurred, so after the second three $H$'s the qubits are in the state
\begin{equation}
\mu|-\alpha\rangle_1|-\alpha\rangle_2|-\alpha\rangle_3+\nu|\alpha\rangle_1|\alpha\rangle_2|\alpha\rangle_3.
\end{equation}
We can compare the states of the first and second modes by sending them into a beam splitter, and then measuring $n_2$ the number of photons in the second mode. If neither mode 1 nor mode 2 has received a $Z$ error, then $n_2=0$, and the system is left in the state
\begin{equation} \mu|-\sqrt{2}\alpha\rangle_{1}|0\rangle_{2}|-\alpha\rangle_{3}+\nu|\sqrt{2}\alpha\rangle_{1}|0\rangle_{2}|\alpha\rangle_{3}.
\end{equation}
Mode 2 can be returned to its original state by sending mode 1 through an even beam splitter with the vacuum in the other input mode. We then make a similar comparison on modes 2 and 3, measuring $n_3=0$ photons. The full qubit can be decoded into mode 1 by mixing modes 1 and 2 in a beam splitter with transmissivity of $\eta=\frac{1}{3}$.

If for example, mode 2 suffers from a $Z$ error during the transmission, the state of the three qubits after the decoding $H$s will be
\begin{equation}
\mu|-\alpha\rangle_1|\alpha\rangle_2|-\alpha\rangle_3+\nu|\alpha\rangle_1|-\alpha\rangle_2|\alpha\rangle_3.
\end{equation}
In this case, during the comparison of modes 1 and 2 we would find $n_2$ to be some random integer given by a Poisson distribution whose mean is $2\alpha^2$, and modes 1 and 3 are left in the state
\begin{equation}
\mu\left(-1\right)^{n_2}|0\rangle_1|-\alpha\rangle_3+\nu|0\rangle_1|\alpha\rangle_3,
\end{equation}
which aside from the overall factor of $-1$ is equivalent to 
\begin{equation}
Z^{n_2}\left(\mu|-\alpha\rangle_3+\nu|\alpha\rangle_3\right),
\end{equation}
where we have simply discarded modes 1 and 2. The qubit can now be reconstructed in mode $3$ by application of $Z$ when $n_2$ is odd. If the error instead occurred in mode $3$, we would find $n_2=0$, and $n_3$ is Poisson distributed about $2\alpha^2$. The qubit is then found in mode $1$, requiring $Z$ when $n_3$ is odd. In this way we can detect a $Z$ error in a single mode, and correct the error. If the probability for an error to occur in each of the modes is $P_e$, then the probability that we can transmit an error free qubit using this procedure is $P_s=1-3P_e^2+2P_e^3$.

This three qubit error correction code can be expanded to increase the probability to successfully transmit a qubit by adding more modes to the encoding \cite{Braunstein}. To protect the encoded qubit from a maximum of $n$ errors requires $2n+1$ encoding bits. We must first prepare $2n$ qubits in the $|0\rangle$ state. Then we apply a CNOT to the qubit we are hoping to protect and each of the $2n$ encoding qubits. To complete the encoding, we perform the Hadamard transformation to all of the $2n+1$ qubits. The probability to transmit a qubit without error is now
\begin{equation}
P_s=\sum_{j=0}^{n}\left(\begin{array}{c} 2n+1 \\ j\end{array}\right) P_e^j\left(1-P_e\right)^{2n+1-j}.
\end{equation}
By increasing the number of qubits used in the encoding $P_s$ can increase arbitrarily close to $1$, provided that $P_e<1/2$. Our simplified method for implementation of the three qubit code using optical coherent states can be easily applied to this larger code with $2n+1$ qubits, replacing each of the CNOTs with a beam splitter. Decoding is accomplished by pairwise comparison of neighboring qubits using a beam splitter and a photon counter.

With these methods we can correct the Pauli $Z$ error effecting the qubits as they are transmitted through a long optical fiber.

\section{Amplitude Restoration}

If we plan to transport a coherent state qubit over a long distance, we must have some way to correct the decrease in the amplitude of the coherent states. This can be accomplished using a slightly modified form of teleportation.

Now let us describe how the teleportation scheme can be used to change the amplitude of the coherent states used to encode a qubit. Suppose we want to change the qubit $|Q_{\pm}(\beta)\rangle$ to $|Q_{\pm}(\alpha)\rangle$. In this case, we can restore the amplitude of the qubit using the teleportation scheme in Fig.~\ref{teleport1} with the Bell state
\begin{equation}
|-\beta\rangle_{2}|-\alpha\rangle_{3} + |\beta\rangle_{2}|\alpha\rangle_{3}.
\end{equation}

This state can be made easily if we have a source of ``cat'' states as described above. We first prepare the cat state
\begin{equation}
|-\frac{\beta}{\cos \theta}\rangle + |\frac{\beta}{\cos \theta}\rangle
\end{equation}
and send it into a beam splitter with transmissivity 
\begin{equation}
T= \cos^2{\theta} = \left(1 + \frac{\alpha^2}{\beta^2}\right)^{-1}. 
\end{equation}
Once we have constructed the necessary Bell state, we proceed with teleportation, exactly as described above and pictured in Fig.~\ref{teleport1} except that $|B_{\pm}\rangle$ is replaced with the new Bell state. Because the amplitude of the Bell state's first mode is tailored to match the qubit, they will experience total interference at the beam splitter.  The teleported qubit will then emerge with an amplitude equal to that of the Bell state's second mode.

Note that this technique can also be applied in the $(0,\alpha)$ encoding using an analogous cat state. We can use this method to increase a qubit's amplitude to prepare it for the error correction code of the previous section. This method also allows us to repair the decreased amplitude of a qubit which has suffered from some absorption.

\begin{figure}
\begin{center}
\includegraphics[bb = 92 578 319 716, clip, width=80mm]{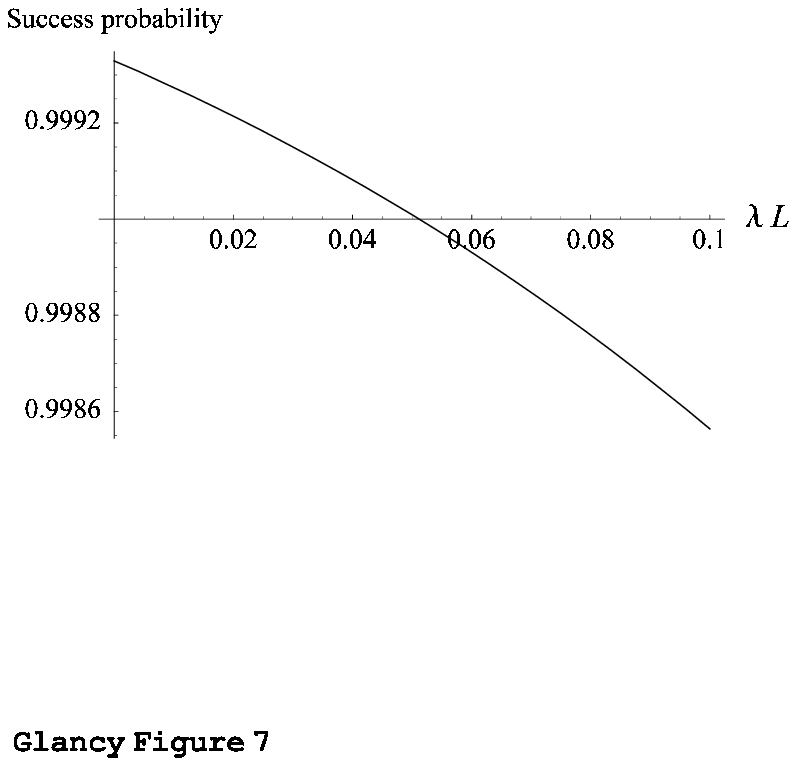}
\end{center}
\caption{\label{fixingsuccess}Teleportation success probability versus $\lambda L$ for $\mu=\nu=\frac{1}{\sqrt{2}}$ and $\alpha = 2$.}
\end{figure}

When we transmit the qubit over a long distance, the coherent amplitude will decrease by the factor $e^{-\lambda L /2}$. Because the teleportation success probability depends on the amplitude $\alpha e^{-\lambda L /2}$ of the coherent state, our ability to correct the amplitude loss will decrease with the distance traveled by the qubit. Fig.~\ref{fixingsuccess} shows the teleportation success probability as a function of the transmitted distance. This tells us when we should perform teleportation to correct the qubit amplitude. 
  
\section{Conclusions}
As a coherent state qubit travels along a fiber-optic cable it suffers from two forms of decoherence. Absorption causes both a decrease in the amplitude of the coherent state and a dephasing in the qubit basis. The amplitude can be restored through teleportation using a specially prepared Bell-state. We can correct the dephasing using the standard three qubit phase-flip correction code. We have also shown how the phase-flip code can be simplified for the coherent state model, and how the phase-flip code can be expanded to provide greater qubit fidelity by using greater numbers of qubits to encode the information.

We might note here that in order to transmit a qubit over a long distance, instead of using an error correcting code as we have described, one could employ quantum teleportation. Suppose Alice wants to transmit her qubit to Bob. Rather than encoding her qubit against $Z$ errors and sending the three qubits to Bob, she could prepare a large collection of Bell states. She then sends one half of each Bell pair to Bob through their long optical fiber. The Bell states will suffer some decoherence during the transmission, so Alice and Bob must purify the states. A method for purification of Bell states (made of coherent states of light) is published in \cite{Jeong2}. A much more thorough discussion of the connections between error correction codes and purification of Bell states can be found in \cite{Bennett}.

Throughout this paper we have also discussed a number of connections between the two methods used to encode qubits onto coherent states: the encoding whose logical $|0\rangle$ and $|1\rangle$ states are the optical coherent states $|-\alpha\rangle$ and $|\alpha\rangle$, and the encoding whose logical states are $|0\rangle$ (the vacuum state) and $|2\alpha\rangle$. We can transform a qubit from one encoding to another by using the displacement operator $D(\pm\alpha)$. Both encodings suffer from the same forms of decoherence at the same rates. The probability to teleport a qubit does not depend on its encoding. In a future publication we hope to discuss the performance of logic operations on both encodings and how the freedom offered by the displacement operator can increase the efficiency of calculations.

In this paper we have not discussed the effects of any optical nonlinearities during transmission of the qubit. However, the nonlinearities of fiber-optic cables at these low light levels are negligible. Of a much greater concern is a need for high quality photon counters and a source of cat states. These are required for the measurements during the error correction and teleportation procedures and for the production of Bell states. Because of the great utility of photon counters and cat states, we encourage experimenters to continue their work on these devices.

\begin{acknowledgments}
We would like to thank Alexei Gilchrist, Gerard Milburn, John LoSecco, and Carol Tanner for helpful discussions. Hilma Vasconcelos thanks the Center for Applied Mathematics at the University of Notre Dame for their financial support.
\end{acknowledgments}

\end{document}